\documentclass[prl,aps,amsmath,amssymb,twocolumn,groupedaddress,floats,final,postprint,superscriptaddress]{revtex4}
\usepackage{graphicx}
\usepackage[caption=false]{subfig}

\usepackage[polutonikogreek,english]{babel}
\usepackage{teubner}
\usepackage{color}
\usepackage{bm}
\usepackage{MnSymbol}%
\usepackage{wasysym}%
\usepackage{algorithm}
\usepackage{hyperref}
\usepackage{float}
\usepackage{pgf}
\usepackage{tikz}
\usepackage{graphicx}

\usepackage[mathlines]{lineno}

\newcommand{\ZZ}{\mathbb{Z}}
\newcommand{\NN}{\mathbb{N}}
\newcommand{\RR}{\mathbb{R}}
\newcommand{\EE}{\mathbb{E}}

\newcommand{\sN}{\mathcal{N}}
\newcommand{\sS}{\mathcal{S}}
\newcommand{\sU}{\mathcal{U}}
\newcommand{\sW}{\mathcal{W}}

\newcommand{\dc}{d_{\mathrm{c}}}

\def\q{{\hbox{\foreignlanguage{greek}{\coppa}}}}
\def\qq{{\hbox{\foreignlanguage{greek}{\footnotesize\coppa}}}}

\begin{document}

\author{Jens Grimm}
\affiliation{ARC Centre of Excellence for Mathematical and Statistical Frontiers
(ACEMS), School of Mathematical Sciences, Monash University, Clayton,
Victoria 3800, Australia}
\author{Eren Metin El\c{c}i}
\affiliation{School of Mathematical Sciences, Monash University, Clayton, Victoria 3800, Australia}
\author{Zongzheng Zhou}
\affiliation{ARC Centre of Excellence for Mathematical and Statistical Frontiers
(ACEMS), School of Mathematical Sciences, Monash University, Clayton,
VIC 3800, Australia}
\author{Timothy M. Garoni}
\affiliation{ARC Centre of Excellence for Mathematical and Statistical Frontiers
(ACEMS), School of Mathematical Sciences, Monash University, Clayton,
VIC 3800, Australia}
\author{Youjin Deng}
\affiliation{Hefei National Laboratory for Physical Sciences at Microscale, Department of Modern Physics, University of Science and Technology of China, Hefei 230027, China and CAS Center for Excellence and Synergetic Innovation Center in Quantum Information and Quantum Physics, University of Science and Technology of China, Hefei, Anhui 230026, China}

\title{Geometric explanation of anomalous finite-size scaling in high dimensions}

\date{\today}

\begin{abstract} 
We give an intuitive geometric explanation for the apparent breakdown of standard finite-size scaling in systems with periodic boundaries
above the upper critical dimension.  The Ising model and self-avoiding walk are simulated on five-dimensional hypercubic lattices with free
and periodic boundary conditions, by using geometric representations and recently introduced Markov-chain Monte Carlo algorithms.  We
show that previously observed anomalous behaviour for correlation functions, measured on the standard Euclidean scale, can be removed by
defining correlation functions on a scale which correctly accounts for windings.
\end{abstract}

\maketitle

\label{sec:introduction}
Finite-size Scaling (FSS) is a fundamental physical theory within statistical mechanics, describing the asymptotic approach to the thermodynamic
limit of finite systems in the neighbourhood of a critical phase transition~\cite{fisher1971,fisher1972}. 

It is well-known~\cite{FernandezFrohlichSokal92} that models of critical phenomena typically possess an upper critical dimension, $\dc$,
such that in dimensions $d\ge\dc$, their thermodynamic behaviour is governed by critical exponents taking simple mean-field
values~\cite{footnote1}.  In contrast to the simplicity of the
thermodynamic behaviour, however, the theory of FSS in dimensions above $\dc$ is surprisingly subtle, and remains the subject of ongoing
debate \cite{flora2016,kenna2016,lundow2016,young2014,lundow2014,kenna2014,kenna2012,lundow2011}. We will show here that such subtleties can
be explained in a simple way, by taking an appropriate geometric perspective.

Perhaps the most important class of models in equilibrium statistical mechanics are the $n$-vector models~\cite{Stanley68}, describing
systems of pairwise-interacting unit-vector spins in $\RR^n$~\cite{footnote2}. The cases $n=1,2,3$ respectively correspond to the Ising,
XY and Heisenberg models of ferromagnetism, while the limiting case $n=0$ corresponds to the Self-avoiding Walk (SAW) model of polymers
~\cite{FernandezFrohlichSokal92}.

The $n$-vector model has wide-ranging applications in condensed matter physics, particularly in the theory of
superfluidity/superconductivity and quantum magnetism. In addition, the case $n=2$ is related to the Bose-Hubbard
model~\cite{SvistunovBabaevProkofev15} which is actively studied in the field of ultra-cold atom physics. In such quantum applications, the
quantum system in $d$ spatial dimensions is related to the classical model in $d+1$ dimensions. Since~\cite{FernandezFrohlichSokal92}
$\dc=4$ for the nearest-neighbour $n$-vector model, this shows that understanding its FSS when $d\ge\dc$ is of importance not only to the
theory of FSS itself, but also in the field of condensed matter physics more generally.  We also note that the value of $\dc$ can be reduced
by the introduction of long-range interactions.

In this Letter, we apply a geometric approach to re-examine a long-standing debate concerning the FSS of the $n$-vector model with $d>\dc$
\cite{flora2016,kenna2016,lundow2016,young2014,lundow2014,kenna2014,kenna2012,lundow2011}. The majority of this debate has focused on the
boundary-dependent FSS of the ferromagnetic Ising model when $d>\dc$; particularly on the case $d=5$.  Numerical
observations~\cite{luijten1999} for the magnetic susceptibility $\chi$ have established an anomalous FSS behaviour $\chi \sim L^{d/2}$ at
the critical point, when using periodic boundary conditions (PBC). By contrast, standard mean field behaviour $\chi \sim L^2$ is
observed~\cite{lundow2014,young2014} for free boundary conditions (FBC). Moreover, with periodic boundaries, it was numerically
observed~\cite{kenna2014} that the correlation between spins at distance $L/2$ scales like $L^{-d/2}$, in contrast to the standard
mean-field prediction $L^{-(d-2)}$ expected for free boundaries.
\begin{figure}[ht]
\includegraphics[scale=0.8]{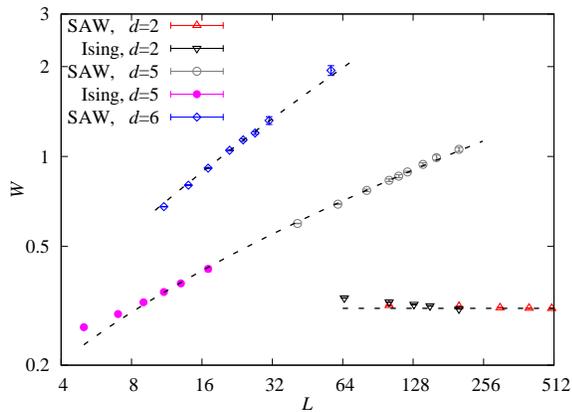}
\caption{Average winding number, $W$, for Ising and SAW models with periodic boundary conditions.
The number of windings is asymptotically constant in $L$ for $d<\dc$. Above $\dc$, windings proliferate with increasing $L$. To emphasize the
universal scaling, the data for Ising and SAW were translated onto a single curve for both $d=2$ and $d=5$.  The predicted scaling $W\sim
L^{\q-1}$ with $\q = d/\dc$ is evident for $d>\dc$.
}
\label{winding}
\end{figure}
Recent renormalization-group arguments attempt to explain this anomalous FSS by postulating a modified scaling of the correlation length $\xi
\sim L^\qq $ when $d>\dc$, where $\q := d/\dc$~\cite{kennareview}. Moreover, an additional exponent $\eta_Q$, related
to $\qq$, was introduced to explain the anomalous large-distance behaviour of the spin-spin correlation function.

Our central message is that in order to elucidate the mechanisms underlying boundary-dependent FSS, it proves useful to consider
appropriate geometric representations of the investigated systems. In particular, rather than working directly with Ising spins, we study
the Ising model via its high-temperature representation~\cite{thompson1979}. In addition, we also study the analogous FSS
properties for the self-avoiding walk.\par
We argue that the apparent breakdown of standard FSS for periodic systems is a manifestation of the proliferation of windings (see below for
a precise definition), absent for dimensions below $\dc$; see Fig.~\ref{winding}. This motivates introducing an alternative definition of
length on the torus, different from the standard Euclidean length, which accounts for the number of such windings.  We refer to this length
scale, which is of order $WL\sim L^{\q}$, as the \emph{unwrapped length}.  Our numerical results show that when correlations are measured on
the scale of the unwrapped length, the apparently anomalous behaviour on the torus disappears, and standard mean-field behaviour is
recovered.

\begin{figure*}[ht!]
\includegraphics[scale=0.7]{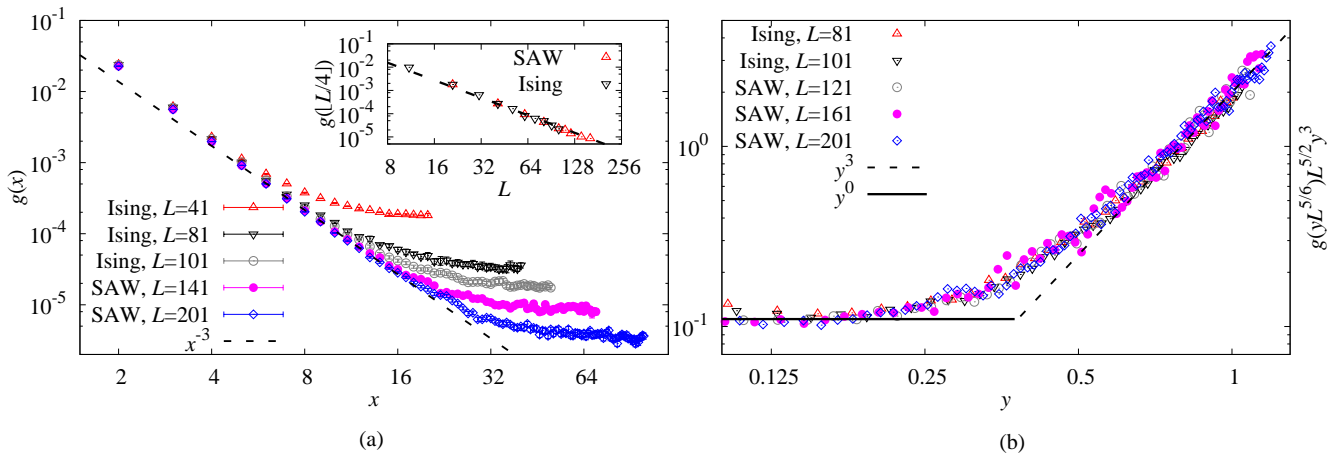}
 \caption{
 (a) Correlation functions for SAW and Ising models with PBC, on the Euclidean scale.
   The inset shows FSS at the fixed point $x=\lfloor L/4 \rfloor$ leading to $g_{\text{Ising+SAW,PBC}}(\lfloor L/4 \rfloor) \sim
   L^{-d/2}$ for both models, in contrast to the standard mean field prediction of $L^{-3}$.
 (b) Collapse of both the Ising and SAW data with $y:= x/L^{d/[2(d-2)]}$ onto the ansatz~\eqref{eq:pap}.}
 \label{correlation_function_pbc}
\end{figure*}

Furthermore, we provide strong numerical evidence that even when measured on the Euclidean scale, the observed anomalous behaviour of
critical correlations on the torus can be explained without introducing a new critical exponent $\eta_Q$.
In particular, Fig.~\ref{correlation_function_pbc}a and~\ref{correlation_function_pbc}b confirm the following piecewise asymptotic behaviour
\begin{equation}
\langle s_0 s_\mathbf{x} \rangle_{PBC} 
\sim 
\begin{cases} \|\mathbf{x}\|^{-(d-2)}, & \|\mathbf{x}\| \le O\left(L^{d/[2(d-2)]}\right),\\ 
L^{-d/2},  & \|\mathbf{x}\| \ge  O\left(L^{d/[2(d-2)]}\right),
\end{cases}
\label{eq:pap}
\end{equation}
as conjectured in \cite{papathanakos2006}. These figures also show that the analogous quantity for SAWs obeys the same scaling.\medskip

\emph{Geometric representations and observables.---} 
The zero-field Ising model is defined by the Hamiltonian $\mathcal{H} = - \sum_{ij} s_i s_j$, where $s_i \in \{-1,+1\}$ is
the spin on site $i$ of a hypercubic lattice of side length $L$, and the sum is over nearest neighbours.
Its so-called high-temperature expansion \cite{thompson1979} provides a natural geometric representation for the spin-spin correlation function
\begin{equation}
g_{\text{Ising}}(\mathbf{x}) := \langle s_0\,s_\mathbf{x}\rangle = \frac{\sum_{A : \partial A=\{0,\mathbf{x}\}} z^{|A|}}{\sum_{A:\partial A = \emptyset} z^{|A|}},
\label{eq:ising_correlation_function}
\end{equation}
where the sums are over all bond configurations $A$ subject to the given constraint on $\partial A$, where $\partial A$ denotes the set of 
vertices incident to an \emph{odd} number of occupied bonds.
The bond fugacity satisfies $z=\tanh(1/T)$, where $T$ is the Ising temperature. 
We simulated this geometric representation using the worm algorithm of Prokof'ev and Svistunov~\cite{prokofev}.

We also considered the SAW model in the grand-canonical (variable-length) ensemble. 
The SAW correlation function is given by
\begin{equation}
g_{\text{SAW}}(\mathbf{x}):=\sum_{\omega\,:\,\ 0 \to \mathbf{x}} z^{|\omega|},
\label{eq:saw_correlation_function}
\end{equation}
where the sum is over all SAWs beginning at the origin and ending at site $\mathbf{x}$.
In contrast to typical studies of SAWs, which are performed on the infinite lattice,
we consider this model to be confined to finite subsets of the hypercubic lattice, as studied for the Ising model.
We simulated this ensemble using an irreversible version
of the Beretti-Sokal (B-S) algorithm \cite{beretti1985} introduced in \cite{deng2017}.

For convenience, we measured the Ising and SAW correlation functions only along the first coordinate axis, i.e. only at
$\mathbf{x}=(x,0,\ldots,0)$ for $0\le x \le L/2$.

In addition, we measured the winding number $\sW$:
\begin{itemize}
\item For the SAW model, $\sW$ is defined as the number of windings along the first coordinate axis.
\item In the Ising model, $\sW$ is defined as the number of windings along the first coordinate axis, in the largest cluster.
Measurements were taken when $\partial A = \emptyset$.
\end{itemize}
We emphasize that $\sW$ does not distinguish between windings in the positive or negative directions, and takes strictly non-negative values.
We denote the mean value of $\sW$ by $W:=\langle \sW\rangle$.

Furthermore, for the SAW model, we additionally measured:
\begin{itemize}
\item The walk length $\sN$, and its mean $N$.
\item The unwrapped correlation function $\tilde{g}_{\text{SAW}}:\NN\to\RR$
\begin{equation}
\tilde{g}_{\text{SAW}}(u) := \sum_{\omega\in\sS_1 \,:\, \sU(\omega) = u} z^{\sN(\omega)}
\label{unwrapped correlation function definition}
\end{equation}
where $\sS_1$ is the set of all SAWs on $\ZZ_L^d$ which start at the origin and end on the first coordinate axis,
and where the unwrapped length $\sU$ is defined algorithmically as follows.
For $\omega\in\sS_1$, traverse $\omega$ from the origin to its endpoint, adding
$+1$ ($-1$) for each step of the walk in the positive (negative) direction along the first coordinate axis.
\end{itemize}
The unwrapped length $\sU$ simply corresponds to the length the walk would have in the infinite lattice, if the torus were unwrapped, so that
periodic images are considered distinct.

Our simulations for both models were performed on the standard hypercubic lattice, using both FBC and PBC.  The Ising model was simulated at the
exact infinite-volume critical point in two dimensions~\cite{baxter1982}, and at the estimated location of the infinite-volume critical
point $z_{c,\text{Ising,5d}} = 0.113~915~0(5)$~\cite{lundow2014} in five dimensions. The SAW model was simulated at the estimated location
of the infinite-volume critical points, $z_{c,\text{SAW,2d}} = 0.379~052~277~758(4)$ \cite{jensen2003}, $z_{c,\text{SAW,5d}} =
0.113~140~84(1)$\cite{deng2017} and $z_{c,\text{SAW,6d}} = 0.091~927~86(4)$ \cite{Owczarek2001}, in dimensions, 2, 5 and 6,
respectively. Our fitting methodology and corresponding error estimation follow standard procedures, see e.g. \cite{young_book,
  sokal1996}. To estimate the exponent value for a generic observable $Y$ we performed least-squares fits to the ansatz $Y = a_Y L^{b_Y} + c_Y$.

To estimate, $g_{\text{Ising, PBC/FBC}}$ we achieved linear system sizes up to $L = 101$. 
Our SAW simulations were performed in the range $21 \le L \le 201$
for PBC and $51 \le L \le 401$ for FBC, respectively. A detailed analysis of autocorrelation times can be found in \cite{deng2007} for the
worm algorithm and in \cite{deng2017} for the irreversible B-S algorithm.\medskip

\begin{figure}[ht]
\includegraphics[scale=0.8]{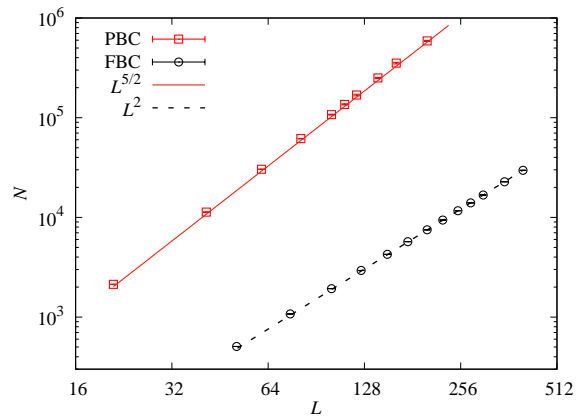}%
\caption{FSS of the average walk length $N$ of a critical SAW in five dimensions. As predicted in the text, on the torus we observe $N\sim L^{d/2}$. 
By contrast, with free boundaries we observe the standard mean-field behaviour $N\sim L^2$. 
}
\label{meanwalklength}
\end{figure}

\emph{Boundary dependent FSS, and choosing the right scale.---}
We now present a scaling argument which characterizes the proliferation of windings in the SAW model in terms of the exponent $\q$.
Consider a uniformly random SAW of fixed length $N$ in $\ZZ^d$, with $d>\dc$. The second virial coefficient $B^{N,N}_2$ provides a measure
of the excluded volume between a pair of such SAWs, and is believed to scale like $B^{N,N}_2\sim N^2$ (see e.g.~\cite{li1994}).  This
suggests that in order to wrap such a walk onto a torus $\ZZ_L^d$, without introducing intersections, would require $N^2\lesssim L^d$.
Considering now a variable length ensemble at $z_c$, we expect the mean of $\mathcal{N}$ to be of the order of its maximum, which implies
$\EE(\mathcal{N})\sim L^{d/2}$. Figure~\ref{meanwalklength} verifies this prediction. 
Furthermore, if one were to take a typical SAW on the torus $\ZZ_L^d$, and unwrap it into $\ZZ^d$, it would have root-mean-square
displacement of order $W L$. But for a uniformly-random fixed-length SAW in $\ZZ^d$ with $d>\dc$, the mean-square displacement scales like
the walk length. Combining this with the above observation shows that $W L \sim L^{d/4} = L^{\qq}$.

Figure~ \ref*{winding} confirms that this prediction holds for the SAW model with $d=5,6$, and also for the Ising model with $d=5$.
By contrast, no proliferation of windings is observed for $d=2$ in either model.
Our fits yield $b_{W,\text{SAW,5d}}=0.27(4)$ and $b_{W,\text{Ising,5d}}=0.24(3)$ and $b_{W,\text{SAW, 6d}}=0.54(7)$, in good agreement with 
the predicted value of $\q-1$.

\medskip

\emph{Unwrapped correlation function.---}
Consider simple random walk (SRW) on a finite torus $\ZZ_L^d$ with $d>2$. If one defines $g_{\mathrm{SRW}}(x)$ analogously
to $g_{\mathrm{SAW}}(x)$, it diverges at the SRW critical fugacity $z=1/2d$. However, if instead one considers the unwrapped version, defined by
summing over simple random walks in~\eqref{unwrapped correlation function definition}, rather than SAWs, then one immediately recovers the
usual infinite-lattice simple-random walk Green's function. 

\begin{figure}[ht!]
\includegraphics[scale=0.8]{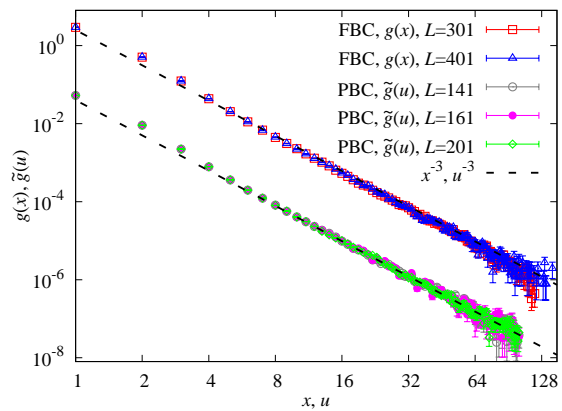}
    \caption{
  Comparison of $g_{\text{SAW, FBC}}$ and $\tilde{g}_{\text{SAW, PBC}}$.
  For clarity, the data for $g_{\text{SAW,FBC}}(x)$ was translated upwards.
}
    \label{corr_fbc_unwrapped}
\end{figure}

This simple observation suggests that the unwrapped SAW correlation function $\tilde{g}_{\mathrm{SAW}}(u)$ may also recover the standard
mean-field behaviour. Figure~\ref{corr_fbc_unwrapped} clearly illustrates that this is indeed the case. From the figure, we see that the
unwrapped correlation function for systems with PBC displays identical scaling behaviour to the Euclidean correlation function for systems
with FBC. Numerically fitting the exponent corresponding to the power-law decay $\tilde{g}_{\mathrm{SAW,PBC}}(u)\sim u^{-b}$ yields
$b=-3.02(2)$. Similarly, fitting the decay exponent for $g_{\mathrm{SAW,FBC}}(x)$ yields $b=-3.01(2)$.
Both estimates are in excellent agreement with the expected $x^{-(d-2)}$ scaling corresponding to the infinite-lattice SRW Green's function.

\emph{Correlations on the Euclidean scale.---}
Finally, we now consider the correlation functions for the Ising and SAW models with periodic boundary conditions, on the Euclidean scale.
As predicted by~\eqref{eq:pap}, 
Fig.~\ref*{correlation_function_pbc}a shows that two qualitatively different regions of $x$ can be identified.
At short distances, both correlation functions exhibit standard mean-field decay $x^{-(d-2)}$, while at long distances, both enter a plateau.
In the inset, we fix a point in the
plateau, $x=\lfloor L/4 \rfloor$, and analyze the $L$ dependence of the correlations at this point. Our fits yield
exponent values of $-2.55(12)$ and $-2.52(23)$ for SAW and Ising, respectively, in good agreement with the prediction $d/2$ from~\eqref{eq:pap}.
This $L^{-d/2}$ scaling in the bulk is in agreement with a previous study of the Ising model presented in~\cite{kenna2014}.\par
To further test the conjectured scaling form~\eqref{eq:pap}, Fig.~\ref{correlation_function_pbc}b plots appropriately scaled versions of
$g_{\mathrm{SAW,PBC}}$ and $g_{\mathrm{Ising,PBC}}$ against the dimensionless variable $y:= x/L^{d/[2(d-2)]}$. The excellent data collapse
provides strong evidence for the validity of~\eqref{eq:pap}.\medskip

\emph{Discussion.---} 
In this Letter, we have studied boundary-dependent FSS above $\dc$ for both the SAW model, and for a geometric representation of the Ising
model. We have established that the anomalous behaviour observed previously for the correlation functions can be explained geometrically,
without any need for new critical exponents.  This conclusion is in broad agreement with independent arguments made previously
in~\cite{young2014}. In that work, an analysis of the Fourier modes of the Ising model was presented, which also refuted the need for the
exponent $\eta_Q$. 

Moreover, our results show that if one considers correlations of the periodic system on the scale of the unwrapped length, rather than the Euclidean length,
then standard mean-field behaviour is recovered. Furthermore, the scale of the unwrapped length is shown to be governed by the exponent $\q$.

Our consideration of unwrapped correlations above focused on the SAW case, largely for reasons of computational efficiency. However we
expect analogous constructions to apply to the Ising case. Unlike the SAW case, there does not appear to be one unique sensible choice for
the definition of unwrapped length in the Ising case. One sensible candidate would appear to be the longest path along one fixed axis
between the odd-degree vertices. Unfortunately, finding the longest path in a graph is a computationally demanding task, which makes
testing this conjecture challenging \cite{schrijver2003}.

While we have focused on the case $d>\dc$, we expect that similar phenomena will be observed also \textit{at} $d=\dc$. In this case,
however, the logarithmic multiplicative corrections to the mean-field thermodynamics will make the analysis more subtle. Nonetheless, our
preliminary simulations of the XY model at $d=\dc$ suggest that the windings, which are directly related to the superfluid density in this
case, are again divergent. This suggests that the correlation function may again exhibit the two-scale behaviour displayed
in~\eqref{eq:pap}. Such phenomena will likely have important consequences in studies of condensed matter in three spatial dimensions, in
particular to quantum critical dynamics~\cite{sandvik2017}. 

Finally, we note that anomalous FSS behaviour on tori has been established rigorously for percolation and the Loop Erased Random Walk
(LERW). For LERW, it has been shown~\cite{benjamini2005} that the mean path length scales as $L^{d/2}$, in agreement with our observations
for SAW. For percolation it was conjectured that for $d>6$ \cite{aizenman1997} the largest cluster scales as $L^4$ for bulk boundary
conditions while it scales as $L^{2d/3}$ with periodic boundaries. This conjecture was subsequently proved, for sufficiently large
dimension, for bulk boundaries in~\cite{hara2008} and for periodic boundaries in~\cite{heydenreich2007}. 
It would be of significant interest to study such percolative questions in the general framework of the random cluster model, and
examine the Ising model from this alternative geometric perspective.

\begin{acknowledgments}
\emph{Acknowledgments.---} We would like to thank S. Bowly, A. Collevecchio, H. Hu, R. Kenna, M. Weigel, U. Wolff and J.-S. Zhang for fruitful
discussions. This work was supported under the Australian Research Council’s Discovery Projects funding scheme (Project Number
DP140100559). It was undertaken with the assistance of resources from the National Computational Infrastructure (NCI), which is supported by
the Australian Government. Furthermore, we would like to acknowledge the Monash eResearch Centre and eSolutions-Research Support Services
through the use of the Monash Campus HPC Cluster. Y. Deng thanks the National Natural Science Foundation of China for their support under
Grant No. 11625522 and the Fundamental Research Funds for the Central Universities under Grant No. 2340000034. Y. Deng also thanks the support from MOST under Grant No. 2016YFA0301600. J. Grimm and E. M. El\c{c}i thank the University of Science and Technology of China for its hospitality during which this work was written.
\end{acknowledgments}

\end{document}